%% file: main.tex
\documentclass[10pt,letterpaper,compsoc,conference]{template/iiswc25}

\usepackage{cite}
\usepackage{amsmath,amssymb,amsfonts}
\usepackage{algorithmic}
\usepackage{graphicx}
\usepackage[dvipsnames]{xcolor}
\usepackage[final]{microtype}
\usepackage[italic]{mathastext}
\usepackage{libertine}
\usepackage[T1]{fontenc}
\usepackage{textcomp}
\usepackage[varqu,varl]{zi4}
\usepackage[all]{nowidow}
\usepackage[auth-lg,affil-it]{authblk}
\usepackage[keeplastbox]{template/flushend}
\usepackage{fancyhdr}

\usepackage{subfigure}
\usepackage{ulem}

\fancypagestyle{firstpage}{
  \fancyhf{}
  
  \fancyfoot[C]{\thepage}
}

\newcommand*{\RELEASE}{} 

\ifdefined\RELEASE
  \newcommand{\ignore}[1]{}
  \newcommand{\fixme}[1]{}
  \newcommand{\dmi}[1]{}
  \newcommand{\leo}[1]{#1}
  \newcommand{\hc}[1]{}
  \newcommand{\lazar}[1]{}
  \newcommand{\bx}[1]{}
  \newcommand{\kev}[1]{}
  \newcommand{\dha}[1]{}
  \newcommand{\col}[1]{}
  \newcommand{\hon}[1]{}
  \newcommand{\TODO}[1]{}

\else
  \newcommand{\ignore}[1]{}
  \newcommand{\fixme}[1]{{\textcolor{red}{[~FIXME:~#1~]}}}
  \newcommand{\dmi}[1]{{\textcolor{blue}{[~D:~#1~]}}}
  \newcommand{\leo}[1]{{\textcolor{teal}{[~Le:~#1~]}}}
  \newcommand{\hc}[1]{{\textcolor{gray}{[~H:~#1~]}}}
  \newcommand{\lazar}[1]{{\textcolor{olive}{[~LA:~#1~]}}}
  \newcommand{\bx}[1]{{\textcolor{brown}{[~B:~#1~]}}}
  \newcommand{\dha}[1]{{\textcolor{orange}{[~Dh:~#1~]}}}
  \newcommand{\col}[1]{{\textcolor{orange}{[~Co:~#1~]}}}
  \newcommand{\hon}[1]{{\textcolor{orange}{[~Ho:~#1~]}}}
  \newcommand{\kev}[1]{{\textcolor{orange}{[~Ke:~#1~]}}}
  \newcommand{\TODO}[1]{{\textcolor{red}{TODO:~#1}}}
  
\fi

\begin{document}


\title{The High Cost of Keeping Warm: Characterizing Overhead in Serverless Autoscaling Policies}





\author[1]{Leonid Kondrashov}
\author[1]{Boxi Zhou}
\author[2]{Hancheng Wang}
\author[1]{Dmitrii Ustiugov}
\affil[1]{NTU, Singapore}
\affil[2]{Nanjing University, China}

\maketitle
\thispagestyle{firstpage}
\pagestyle{plain}


\input{sections/0-abstract}


\input{sections/1-intro}

\input{sections/2-background}
\input{sections/3-methodology}

\input{sections/4-characterization}

\input{sections/5-discussion}

\input{sections/6-related}

\input{sections/7-conclusion}

\bibliographystyle{template/IEEEtranS}

\bibliography{./bibcloud/gen-abbrev,ref,dblp}

\end{document}

%% file: sections/0-abstract.tex
\begin{abstract}

Serverless computing is transforming cloud application development, but the performance-cost trade-offs of control plane designs remain poorly understood due to a lack of open, cross-platform benchmarks and detailed system analyses. In this work, we address these gaps by designing a serverless system that approximates the scaling behaviors of commercial providers, including AWS Lambda and Google Cloud Run. We systematically compare the performance and cost-efficiency of both synchronous and asynchronous autoscaling policies by replaying real-world workloads and varying key autoscaling parameters.

We demonstrate that our open-source systems can closely replicate the operational characteristics of commercial platforms, enabling reproducible and transparent experimentation. By evaluating how autoscaling parameters affect latency, memory usage, and CPU overhead, we reveal several key findings. First, we find that serverless systems exhibit significant computational overhead due to instance churn equivalent to 10–40\% of the CPU cycles spent on request handling, primarily originating from worker nodes.
Second, we observe \leo{high memory allocation due to scaling policy:} 2–10 times more than actively used.
Finally, we demonstrate that reducing these overheads typically results in significant performance degradation in the current systems, underscoring the need for new, cost-efficient autoscaling strategies.
Additionally, we employ a hybrid methodology that combines real control plane deployments with large-scale simulation to extend our evaluation closer to a production scale, thereby bridging the gap between small research clusters and real-world environments. 

\end{abstract}

%% file: sections/1-intro.tex
\section{Introduction}









Serverless computing has rapidly emerged as a foundational paradigm for modern cloud applications. It enables developers to focus on the application logic by automatically handling infrastructure management, including scaling, monitoring, and provisioning. Due to these advantages, numerous serverless platforms such as AWS Lambda~\cite{aws-lambda-scaling}, Google Cloud Run~\cite{gcr}, and Azure Functions~\cite{Azure_functions} are widely adopted and collectively process billions of invocations daily. However, as organizations increasingly rely on serverless architectures to power latency-sensitive and mission-critical workloads, performance bottlenecks and resource inefficiencies in these systems can significantly impact both user experience and operational costs. Therefore, understanding and optimizing the performance-cost trade-offs in serverless systems has become increasingly important.

However, understanding and optimizing the performance-cost trade-offs in serverless systems faces several challenges: (1) The lack of open-source systems that can replicate the behaviors of commercial platforms prevents researchers from thoroughly analyzing these platforms' internal scaling mechanisms. (2) The absence of studies comparing scaling policies across different providers under controlled conditions prevents understanding of optimal scaling strategies for different workload characteristics. (3) The lack of detailed analysis of performance-cost trade-offs under realistic scaling policies prevents providing effective guidance for practical system optimizations. (4) The lack of large-scale system evaluation prevents the discovery of real performance bottlenecks in production environments.

In this work, we address the four challenges mentioned above by designing a serverless system that approximates the scaling behaviors of commercial providers. To our knowledge, we are the first to approximate the scaling behaviors of real-world providers such as AWS Lambda and Google Cloud Run. Through evaluation using real production workloads, our study reveals several key findings:

First, we find that serverless systems \leo{creates noticeable} computational overhead (10-40\% higher CPU utilization) \leo{due to instance churn}, primarily originating from worker nodes. \leo{This indicates that the sandbox management and resource allocation are major sources of overheads and} that optimization efforts should focus on those components rather than the control plane. Second, we observe severe memory overprovisioning (2-10$\times$ more than actively used) \leo{caused by autoscaling policy}. \leo{That means that future works should focus on improving the scaling policy impact on memory consumption} rather than per-sandbox overheads. Finally, we find that while methods exist to reduce \leo{operational costs, they lead to degraded performance, prompting for a creation of more resource-efficient policies.}

\leo{Additionally, we introduce a simulation methodology that can expand the evaluation scale in research settings. We employ KWOK to simulate worker nodes, while retaining the real control plane components. With this methodology, we validate our findings at a scale more representative of a real cloud provider.}

Our main contributions are:

\begin{itemize}
    \item We design an open-source, scalable serverless system based on Knative that approximates the scaling behavior of commercial providers.\footnote{We will release the code, traces, and toolchain after publication.}
    
    \item 
    We characterize how autoscaling parameters affect end-to-end performance, memory usage, and CPU overhead in synchronous and asynchronous policies. We show that parameter settings that improve performance come with significant overhead: memory overprovisioning (2-10$\times$ more than actively used) and computational overhead (10-40\% of useful work).
    
    \item We provide a comparison of synchronous and asynchronous policies in the memory-performance trade-off space, revealing which parameter adjustments are most cost-efficient for achieving performance targets.
    
    \item We validate our findings at a large scale, confirming that the same performance-cost trade-offs hold in production-scale environments.
\end{itemize}

%% file: sections/2-background.tex
\section{Background}











\subsection{Serverless Autoscaling Operation}


\begin{figure}
    \centering
    \subfigure[Warm Start.]{\includegraphics[width=0.7\linewidth]{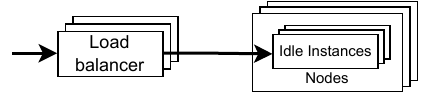}\label{fig:autoscaling:warm}}
    \subfigure[Cold Start (Synchronous).]{\includegraphics[width=0.9\linewidth]{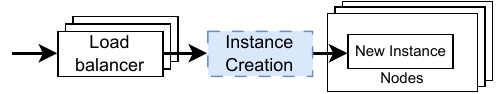}\label{fig:autoscaling:cold_sync}} 
    \subfigure[Cold Start (Asynchronous).]{\includegraphics[width=0.9\linewidth]{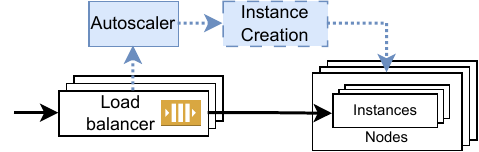}\label{fig:autoscaling:cold_async}} 
    \caption{High-level overview of synchronous and asynchronous autoscaling. 
    }
    \label{fig:autoscaling}
\end{figure}











The overall performance and cost of the serverless system depend on the efficiency of the underlying autoscaling infrastructure, which aims to tailor the number of function instances to the changes in the function's invocation traffic. Specifically,
autoscaling systems are fundamental to modern serverless platforms, enabling dynamic resource allocation in response to fluctuating workloads. Previous research shows that changes in the autoscaling policy can improve the cold start rate~\cite{singhvi:atoll} and cost~\cite{shahrad:serverless, roy:icebreaker}.

Figure~\ref{fig:autoscaling} illustrates the primary request handling pathways and scaling strategies employed in such systems.
Figure~\ref{fig:autoscaling:warm} shows the system workflow when handling a request in the basic case.
When a new request arrives, the load balancer initially attempts to route it to an available idle instance -- the sandbox with user code that would handle the request. This scenario, referred to as a \emph{warm start}, incurs no initialization overhead and is functionally equivalent to the request routing observed in conventional microservice architecture, such as the applications in DeathStarBench~\cite{yu:deathstarbench}.

In contrast, if no idle instances are available, the system must provision new resources, resulting in a \emph{cold start} penalty. Figures~\ref{fig:autoscaling:cold_sync} and \ref{fig:autoscaling:cold_async}
distinguish between the two principal approaches to handling cold start scenarios: synchronous and asynchronous~\cite{liu:gap}, described in \S\ref{sec:sync} and \S\ref{sec:async}, respectively.

We describe in detail the differences between these approaches, as they can significantly impact how these systems handle request queueing and instance creation. Existing works~\cite{cvetkovic:dirigent, cvetkovic:neglected} show that the majority of cold start delays originate from control plane delays.


\subsubsection{Synchronous Approach}
\label{sec:sync}

Some providers (e.g., AWS Lambda~\cite{aws-lambda-scaling}) and open-source systems (e.g., OpenWhisk~\cite{openwhisk}) employ a synchronous approach for scaling.
In the synchronous approach (Figure~\ref{fig:autoscaling:cold_sync}), the load balancer buffers the incoming request until a new instance is fully initialized and ready to process it.
With this approach, the system reacts to traffic changes immediately upon arrival, requesting a new instance when there are not enough instances to handle the invocation.
At the same time, it places the instance creation delay directly on the critical path of invocation handling, which can contribute significantly to the end-to-end latency of the service.

For instance, the simplest synchronous policy that is widely used as a baseline would be the fixed keepalive policy~\cite{shahrad:serverless, singhvi:atoll, roy:icebreaker}. 
In this policy, instance creation occurs on the critical path of invocation handling. The created instances are kept in the idle state for a fixed period of inactivity, called the keepalive period.

\subsubsection{Asynchronous Approach}
\label{sec:async}

Google Cloud Run~\cite{knative_offerings} and Knative systems~\cite{knative-autoscaling} use another, asynchronous approach.
This approach, depicted in Figure~\ref{fig:autoscaling:cold_async}, decouples request buffering from instance provisioning. Here, incoming requests are queued until any instance -- either newly created or one that has completed a prior execution -- becomes available. Instance scaling is managed by a dedicated autoscaler component, which continuously monitors system load and can employ predictive models to anticipate demand and preemptively provision resources. In such a system, the instance creation delay is not on the critical path; however, it can influence the amount of queuing \leo{requests in the load balancer} that occurs during function invocation traffic spikes.



Knative, as a popular serverless system with asynchronous autoscaling, can demonstrate the key difference in the approach. Knative deploys a separate service, Autoscaler, which is responsible for allocating resources to the functions. The autoscaler collects concurrency metrics and makes a scaling decision based on the critical path of execution~\cite{knative-flow}.

\subsection{Trade-Offs in Serverless Autoscaling}












To remain competitive, serverless systems must carefully balance performance, operational costs (including memory and compute resource usage), and security or isolation guarantees. Achieving optimal trade-offs among these often conflicting objectives is a central challenge in designing serverless systems.

\noindent\textbf{Memory-Performance Trade-Off.}
A major performance concern in serverless systems is the cold start latency, which occurs when a function instance must be initialized to handle a request. Cold starts can significantly delay response times and may result in service-level objective (SLO) violations. To mitigate the cold start penalty, platforms may cache warm instances in memory~\cite{shahrad:serverless} or proactively create instances in anticipation of the traffic~\cite{roy:icebreaker, singhvi:atoll}, enabling faster request handling and improved performance. However, this approach increases memory usage, which raises operational costs~\cite{agache:firecracker} and potentially limits scalability due to resource constraints. Thus, system designers must carefully balance the benefits of reduced cold start latency against the increased memory usage required to maintain cached instances.

Cloud service providers can adjust the parameters of each scaling policy to achieve high-level goals, such as a performance-oriented system, a cost-optimized system, or a balance point that strikes an intermediate balance.
For example, a fixed keepalive policy (\S\ref{sec:sync}) can increase the keepalive period to exchange the additional memory costs for fewer cold starts~\cite{shahrad:serverless}. Knative autoscaling also has parameters that affect its behavior, including the history window size and per-instance utilization target.\footnote{Knative's autoscaler component maintains a history of request concurrency data. It then computes the average request concurrency over the history window. It divides it by the per-instance utilization target to determine the desired number of instances, which are then provisioned through Kubernetes.}
A longer history window makes the policy more inertial, while the per-instance utilization target is used to allocate more instances to handle traffic fluctuations without performance degradation.



\noindent\textbf{Isolation-Management Overhead Trade-off.}
Isolation is a key security property in multi-tenant serverless environments, but it often comes at the cost of higher management overhead and lower resource utilization. For example, AWS Lambda enforces strong isolation by restricting each function instance to a single concurrent request~\cite{aws-lambda-scaling}. 
This approach minimizes the risk of cross-request interference but can lead to inefficient resource utilization, especially for I/O-bound workloads that do not fully utilize the allocated compute resources~\cite{kaffes:hermod}.
In contrast, Google Cloud Run allows users to configure the concurrency limit per instance~\cite{gcr-scaling}, enabling multiple requests to be served by a single instance. This design reduces the number of instances required, lowers management overhead, and improves resource utilization, but it weakens isolation guarantees. \leo{Management overhead mentioned here is a direct contribution towards operational cost since all CPU computing requires electricity and cooling.}
The choice of concurrency limit thus reflects a trade-off between strong isolation and efficient resource management.

\subsection{Research Gaps in Serverless Control Plane Benchmarking}













Despite the rapid adoption of serverless computing in production settings, it is known for its unreliable performance~\cite{ustiugov:stellar}. Existing works attribute these performance issues to the control plane overheads~\cite{cvetkovic:dirigent, liu:gap}.
However, modern serverless research sheds little light on the requirements and problems that exist in the production environment. We identify the current limitations of serverless research that hinder the improvements of real production systems.



\noindent\textbf{Lack of Provider-Specific Open-Source Systems.}
Current serverless research lacks validated representative and configurable frameworks that can replicate the behaviors of commercial providers, such as AWS Lambda~\cite{awslambda}, Azure Functions~\cite{Azure_functions}, or Google Cloud Functions~\cite{gcr}, to characterize problems that are relevant to the industry.
While frameworks like Apache OpenWhisk~\cite{openwhisk} and OpenFaaS~\cite{openfaas} exist, they implement their policies rather than mimicking specific provider behaviors. An exception to this rule is Knative~\cite{knative}, which is utilized by providers such as Google Cloud Run~\cite{knative_offerings}. However, there is no open-source system available that aims to replicate the behavior of other major providers, such as AWS Lambda and Azure Functions.
This problem forces researchers to experiment on proprietary platforms with limited internal visibility or rely on alternatives that may not reflect production challenges.

\noindent\textbf{Absence of Cross-Provider Policy Comparisons.}
To make relevant research, researchers should be able to understand the scheduling policies and their impact on performance and costs in production settings. At the same time,
serverless research lacks comprehensive studies directly comparing scaling policies and resource management strategies across providers under controlled settings. 
While performance comparisons exist~\cite{ustiugov:stellar, yu:characterizing, kim:functionbench}, they focus on application-level metrics rather than analyzing underlying policy mechanisms. Providers implement different auto-scaling, concurrency limits, and keepalive policies; however, systematic policy comparisons in isolation remain rare, which prevents an understanding of optimal scaling strategies for different workload characteristics. It also conceals the crucial parameters of the systems that impact cost. And because of this, researchers rely on insights occasionally shared by industry partners in conference talks and blog posts to guide their research.

\noindent\textbf{Limited Cost-Performance Trade-off Analysis.}
Improving resource efficiency is a primary goal for serverless platforms, as it directly reduces operational costs and lowers the environmental impact of cloud computing~\cite{shahrad:serverless, roy:icebreaker}. 
Despite that, 
detailed characterization of cost-performance trade-offs with realistic scaling policies remains insufficient. Some works~\cite{shahrad:serverless, roy:icebreaker} have done characterization of such a trade-off, but they use cold start or warm start rate as their proxy for performance. Unfortunately, those metrics have limited applicability when comparing radically different systems (e.g., synchronous vs asynchronous systems due to their different cold start penalty).

\noindent\textbf{Insufficient Large-Scale Control Plane Evaluation.}
To improve the realistic serverless systems, researchers need to evaluate their approaches at scales comparable to production systems. Those systems are estimated to experience peaks of 2000 VM creation requests per second~\cite{cvetkovic:dirigent}.
However, most serverless research operates at scales significantly smaller than those of production systems~\cite{liu:gap}, which severely limits the scope of research problems studied in these settings.

%% file: sections/3-methodology.tex
\section{Methodology}




We aim to systematically compare the performance and cost-efficiency of serverless platforms under realistic workloads. To achieve this, we evaluate four representative systems: two Knative-based designs and two commercial platforms (AWS Lambda and Google Cloud Run). By replaying real-world serverless workloads with In-Vitro~\cite{ustiugov:enabling}, we analyze how different autoscaling strategies affect latency, resource usage, and cost.


\subsection{Evaluated Systems} 


We evaluate four serverless platforms - two open-source systems with configurable control planes and two commercial offerings with fully managed infrastructures. 

\noindent\textbf{Knative-based Systems.} Knative\cite{knative} is a Kubernetes-based serverless platform widely used in commercial offerings~\cite{knative_offerings}. 

Our first system, vanilla Knative (\textbf{Kn}), utilizes an asynchronous control plane that periodically adjusts instance counts based on observed concurrency levels. 
We evaluate this system under a range of autoscaling configurations. Specifically, we vary the per-instance \emph{utilization target} (referred to as container concurrency target in Knative) using values of 0.5 (performance-optimized), 0.7 (default), and 1.0 (cost-optimized).
Additionally, we evaluate the effect of per-instance concurrency limit (referred to as \emph{container concurrency} in Knative).
This parameter caps the number of requests an instance can handle simultaneously. We perform this experiment under the fixed per-instance utilization target of 0.7, varying the concurrency limit across 1, 2, and 4.
Additionally, we vary the autoscaling \emph{window size} from 30 seconds to 1800 seconds to explore responsiveness under different observation granularities.


The second system is Knative-Synchronous (\textbf{Kn-Sync}), a modified version of Knative that employs synchronous instance creation, similar to AWS Lambda. Specifically, we disable Knative's default concurrency-based autoscaling policy and instead modify the autoscaler to immediately create a new instance when no idle instance is available to serve an incoming request.
To mimic AWS Lambda's behavior, we implement an idle instance retention policy. The system proactively retains idle instances for a configurable keepalive period to delay tearing down.
We evaluate Kn-Sync under various \emph{keepalive durations}, ranging from 30 seconds to 
1800 seconds.

\noindent\textbf{Commercial Serverless Platforms.} The remaining two platforms, \textbf{AWS Lambda} and \textbf{Google Cloud Run (GCR)}, are widely used commercial serverless offerings that represent contrasting autoscaling strategies. AWS Lambda adopts a synchronous scaling strategy, where a new instance is launched immediately upon receiving a request if no idle instance is available. In contrast, GCR employs an asynchronous scaling strategy similar to vanilla Knative, adjusting the number of instances periodically based on observed concurrency history and request volume to balance performance and cost.
We leverage the In-Vitro~\cite{ustiugov:benchmarking} framework to replay the same request traces as in our Knative experiments. 
This ensures consistent workload generation across systems and enables the collection of key performance metrics.


\subsection{Hardware and Software Setup in the Local Deployment} 
We run all the Knative-based systems on an 8-node c220g5 Cloudlab cluster~\cite{CloudLab}. Each node has two Intel Xeon Silver 4114 CPUs @ 2.20GHz, each with 10 physical cores, 192~GB DRAM, and an Intel SSD. We disable SMT and fix the CPU frequency to the base frequency for measurement stability.


For Knative-based experiments, we use vHive~\cite{ustiugov:benchmarking}, i.e., Knative v1.13 running on top of Kubernetes v1.29 with containerd v1.6.18.
All container images are assumed to be cached in memory on each node, consistent with prior studies~\cite{cvetkovic:dirigent}. Knative’s panic mode is disabled, and the Activator (Knative’s load balancer) is configured with a single replica to avoid coordination overhead. 
We carefully tune the system for high performance and to avoid control plane bottlenecks. In particular, we increase the request concurrency limit to 10000 in the Kubernetes API server, request rate limit to 1000 requests per second, and request concurrency limit to 1000 in the Controller Manager. We raise the CPU and memory quotas allocated to Knative core components to 28~vCPUs and 30~GB, respectively.



\subsection{Cloud Provider Setups} 
To minimize network-induced latency in cloud provider experiments, we collocate the In-Vitro load generator~\cite{ustiugov:enabling} with the deployed functions.  
For the GCR experiments, we deploy functions in the \texttt{us-central1} region and provision an \texttt{e2-medium} VM in the same region to run In-Vitro for trace replay and metric collection.  
Similarly, for the AWS Lambda experiments, we deploy functions in the \texttt{us-east-1} region and use a \texttt{t2.medium} EC2 instance in the same region to host the In-Vitro client.



\subsection{\emph{Real} Control Plane Deployment with an \emph{Simulated} Large-Scale Cluster}
\label{sec:kwok}


To overcome the prohibitive costs and practical constraints of physical infrastructure in research settings, we use KWOK~\cite{kwok} v0.6.1 to simulate the behavior of 50 worker nodes while retaining a fully operational \emph{real-world, non-simulated} Knative-Kubernetes control plane.
Hence, our method can reduce the cluster size in the experiments by a factor of 10 (from 50 nodes to 5, in our setup) without compromising accuracy and representativeness in control plane performance analysis.

To make the system compatible with simulated nodes,
we modify Knative's Activator: we implement re-routing to a mock service that handles the requests instead of the KWOK-controlled virtual pods. At the same time, the rest of the control plane is unmodified and operates as in a real large-scale deployment.
This hybrid approach is useful for enabling production-scale evaluations within resource-constrained research environments. By modeling node/pod dynamics and stress-testing control plane behavior under scenarios such as burst sandbox creation or tunable instance launch latencies—conditions otherwise unobservable on physical clusters—we faithfully replicate large-scale operational challenges in the control plane without requiring hardware. The methodology preserves authentic control plane interactions and decision-making processes, mirroring real-world deployments, while abstracting worker node workloads computationally. 

This methodology bridges the gap between small-scale experimentation and production-grade system analysis, enabling researchers to rigorously evaluate autoscaling policies, failure modes, and performance boundaries at cost-prohibitive scales within a cost-effective and reproducible framework.


\subsection{Workload}
We use In-Vitro~\cite{ustiugov:enabling} methodology for representative sampling of Azure Functions Trace~\cite{shahrad:serverless} and load generation.
We use a 400-function trace sample that contains 300k invocations over 80 minutes in our experiments.
This specific size of function trace sample was chosen as the biggest load without reaching the computational limit on the worker nodes during the experiment. We empirically determine this threshold by gradually increasing the number of sampled functions and monitoring cluster-wide CPU utilization. The selected trace produces a maximum cluster CPU utilization of less than 100\%, ensuring realistic stress without performance degradation due to CPU contention.
We choose an 80-minute trace duration and discard the first 40 minutes of each run as warm-up, leaving a 40-minute window for stable measurement. This period is sufficiently long to cover multiple autoscaling cycles and observe system behavior under realistic workloads.
We use a synthetic CPU-intensive spin-loop function with durations drawn from the trace, as is done in prior work~\cite{cvetkovic:dirigent, ustiugov:enabling}.


\subsection{Metrics Used for Evaluation}

We use several metrics to evaluate the systems' performance, memory usage, and operational overheads.

\noindent\textbf{End-to-End Performance.}  
We use the geometric mean of the tail (99\textsuperscript{-th} percentiles) of per-function slowdown as the measurement of responsiveness of the system.\footnote{We first calculate per-invocation slowdown by dividing the end-to-end response time by expected execution duration. Then we compute the per-function 99\textsuperscript{-th} percentile of slowdown. Finally, we aggregate per-function slowdown with the geometric mean.} For this metric, the value of 1 would mean the same behavior as in an unloaded system without any delay incurred by additional data plane components (e.g., frontend, validation, authentication).

\noindent\textbf{Memory Usage.} 
To estimate the cost, we use the total memory footprint of all instances in the cluster, normalized to the total memory footprint of non-idle instances. We refer to this metric as \emph{normalized memory usage}. 
This metric reflects the amount of memory allocated to all instances compared to the memory used by instances handling the user invocation.
This metric targets only the scaling policy efficiency, ignoring the possible per-sandbox memory overheads.

\noindent\textbf{System Management Overheads.}  
We measure two additional sources of operational cost: instance creation rate and computational overhead (normalized CPU overhead) of the system. 
The former reflects the intensity of control plane operations and represents both creation and tear-down events.\footnote{The creation and tear-down events are balanced throughout the experiment, as the instance number doesn't exhibit an overall trend for increase or decrease.} The latter captures the CPU cycles consumed by system components \leo{normalized to CPU utilization of user functions.}

We collect these metrics cluster-wide using Prometheus~\cite{prometheus} and the Kubernetes Metrics Server~\cite{k8s-metrics}, ensuring visibility into both workload-facing and infrastructure-level resource consumption.

%% file: sections/4-characterization.tex
\section{Performance Analysis of the Control Plane Designs in Commercial and Open-Source Clouds}

In this section, we evaluate several systems under a realistic workload to answer the following questions:


\begin{itemize}
    \item Can modified open-source systems approximate production serverless systems? (\S\ref{sec:real_provider})
    \item How do the performance, memory usage, and CPU overhead depend on the parameters of synchronous and asynchronous autoscaling policies? (\S\ref{sec:trends})
    \item How do synchronous and asynchronous policies compare to each other in the memory-performance trade-off space? (\S\ref{sec:trade-off})
    \item Can we increase the scale of evaluated systems using simulation techniques to obtain insights from a small, research cluster that can generalize to large-scale production setups? (\S\ref{sec:big_trade-off})
\end{itemize}

\subsection{Approximating Production Systems}
\label{sec:real_provider}





We compare our open-source systems with the real providers (Google Cloud Run~\cite{gcr} and AWS Lambda~\cite{aws-lambda-scaling}) to provide the community with systems that can reproduce the general behavior of real providers.

\subsubsection{Approximating Synchronous System}


\begin{figure}
    \centering
    \subfigure[AWS Lambda]{
    \includegraphics[height=0.55\linewidth]{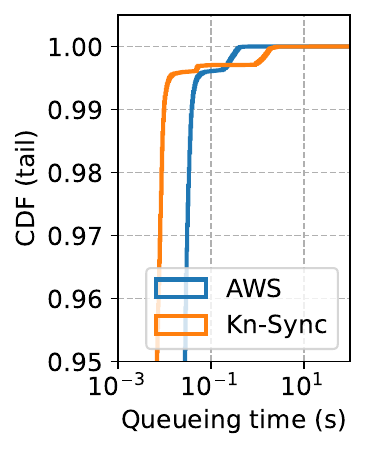}
    \label{fig:aws}
    }
    \subfigure[Google Cloud Run]{
    \includegraphics[height=0.55\linewidth]{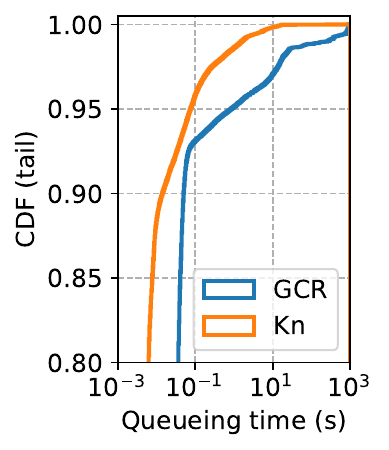}
    \label{fig:gcr}
    }
    \caption{Queueing time in AWS Lambda and Google Cloud Run compared to our Kn-Sync and Kn systems under a realistic workload.}
\end{figure}

We demonstrate that our proposed synchronous system can approximate the behavior of a real provider with a synchronous control plane, such as AWS Lambda~\cite{aws-lambda-scaling}.
%
%
We compare the AWS Lambda performance with Knative-Synchronous, setting the keepalive to 10 minutes. The choice of this specific period is based on the existing evaluation of real providers~\cite{ustiugov:stellar}. Figure~\ref{fig:aws} shows the cumulative distribution function (CDF) of queueing time (time spent in the system without being executed) of those two systems. We focus on worst cases since that's where the scaling policy would have a pronounced impact. Both of them exhibit bimodal behavior, caused by the synchronous system having either no additional delay on warm start or a full instance creation delay on the critical path of invocation. The proportion of cold starts in the measurement is approximately 0.5\% of all invocations for both systems, indicating that the chosen 10-minute keepalive period is close to what AWS Lambda uses. We observe a noticeable difference in performance between the systems for different durations of warm and cold starts. Warm starts in AWS Lambda take 20-30ms, while the Knative-Synchronous achieves warm starts of 5-10ms; this difference indicates a more complex frontend system in a production environment. The difference in cold start durations (approximately 300 ms for AWS and 1 second for Knative-Synchronous) is a result of the use of different sandboxing technologies and optimizations, such as the pooling of pre-booted Firecracker VMs and network namespaces. These results demonstrate that AWS Lambda places instance creations on the critical path of invocations, which take around 300ms. This scaling behavior can be approximated by our Knative-Synchronous system using a fixed keepalive policy with a 10-minute duration.


\subsubsection{Approximating Asynchronous System}



We demonstrate that our asynchronous system can approximate the behavior of a real provider with an asynchronous control plane, such as Google Cloud Run (GCR)~\cite{gcr}.
We compare the performance of GCR with Knative, using a 10-minute history window size.
%
%
Figure~\ref{fig:gcr} shows the CDF tail of per-invocation queueing delays in both of the systems. Here we see that they don't exhibit clear bimodal behavior, but rather demonstrate a linear tail behavior in the queueing time distribution, indicating consistent but variable delays in request processing. We attribute these delays to scaling policies used in GCR and Knative. Knative's policy, based on a window average, may respond too late to traffic changes, leading to additional queueing compared to the fast reaction of a synchronous system, as demonstrated by Knative-Synchronous and AWS Lambda. These results demonstrate that GCR creates a noticeable tail in response time due to its scaling policy, and this scaling behavior can be replicated with the Knative system using a 10-minute window.

\subsection{Performance and Cost Dependency on Parameters}
\label{sec:trends}



We begin our evaluation of the systems by characterizing the impact of autoscalers' parameters on performance and operational costs separately.

\subsubsection{End-to-End Performance}


\begin{figure}
    \centering
    \subfigure[Kn-Sync.]{\includegraphics[width=0.48\linewidth]{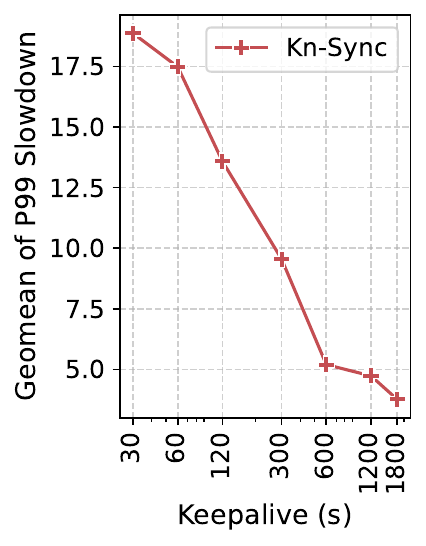}\label{fig:slowdown:sync}} 
    \subfigure[Kn with different utilization targets.]{\includegraphics[width=0.48\linewidth]{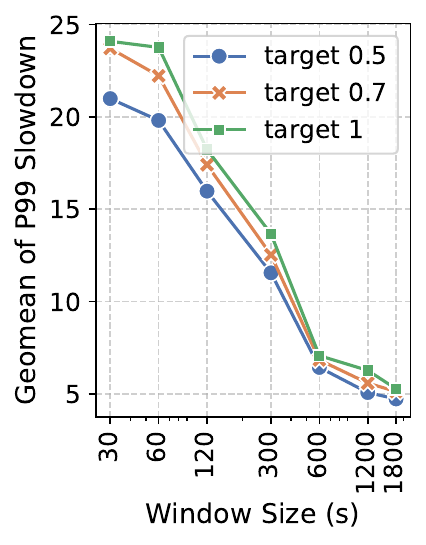}\label{fig:slowdown:async}}
    \caption{Comparison of aggregated slowdowns in Kn-Sync and Kn under a realistic workload with different system parameters.
    }
    \label{fig:slowdown}
\end{figure}



First, we evaluate the end-to-end performance of the evaluated systems.
For the evaluated synchronous system,
we vary the keepalive period from 30 seconds to 1800 seconds and plot the end-to-end worst-case performance in Figure~\ref{fig:slowdown:sync}. The slowdown decreases gradually from 18.9 to 3.8 over the parameter range. Performance hits saturation point at a 600-second keepalive period.

For Knative, we vary two parameters: window size
and per-instance utilization target. We plot the results in Figure~\ref{fig:slowdown:async}. For each target value, we observe a similar trend of reduced slowdown with larger history windows. They all hit the saturation point at a 600-second window (6.4, 6.8, and 7.1 for target values 0.5, 0.7, and 1, respectively). The utilization target affects performance, but on a much smaller scale. For example, at a 600-second window, the slowdown difference between target values of 0.5 and 1.0 is only 9\%.

\textbf{Observation:} \textit{Increasing the keepalive duration or autoscaling window size improves system performance but yields diminishing returns beyond 600 seconds.
A smaller utilization target also results in better performance, although the impact is less significant.}









\subsubsection{Memory Usage}


\begin{figure}
    \centering
    \subfigure[Kn-Sync.]{\includegraphics[width=0.48\linewidth]{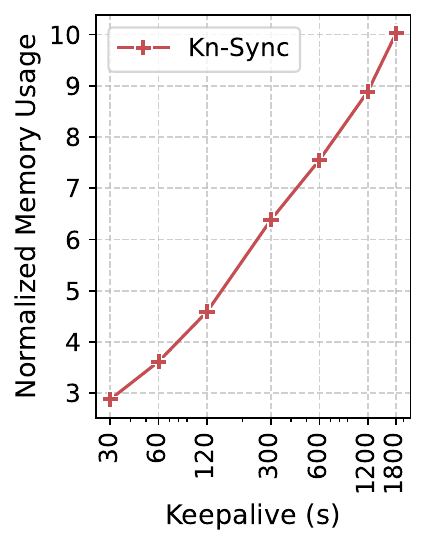}\label{fig:mem_usage:sync}} 
    \subfigure[Kn with different utilization targets.]{\includegraphics[width=0.48\linewidth]{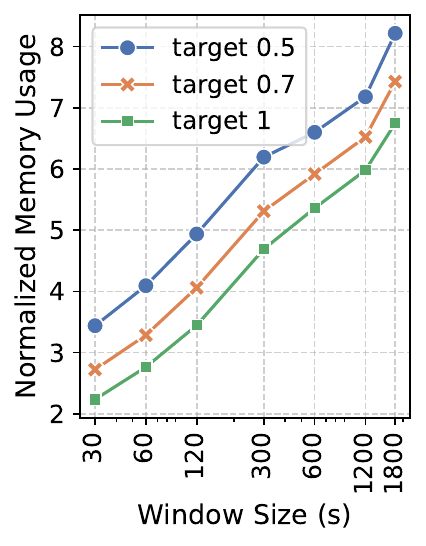}\label{fig:mem_usage:async}} 
    \caption{Comparison of normalized memory usage in Kn-Sync and Kn under a realistic workload with different system parameters.
    }
    \label{fig:mem_usage}
\end{figure}


Next, we evaluate the impact on memory usage.

In Knative-Synchronous, memory usage increases substantially as we increase the keepalive duration. We plot the normalized memory usage in Figure~\ref{fig:mem_usage:sync}. We observe that the normalized memory usage increases from 2.9 to 10 when the keepalive is increased from 30 seconds to 1800 seconds.

In Knative, memory usage increases with larger autoscaling windows and smaller utilization targets. We plot the results in Figure~\ref{fig:mem_usage:async}. When the utilization target is fixed at 0.7, increasing the window size from 30 seconds to 1800 seconds raises the normalized memory usage from 2.7 to 7.4. When the window size is fixed at 600 seconds, decreasing the target value from 1.0 to 0.5 increases memory usage from 5.4 to 6.6. Notably, the effect of reducing the utilization target is less pronounced than that of increasing the window size.



The total memory footprint significantly exceeds the amount required for executing active function instances, reaching 2$\times$ to 10$\times$ the memory needed for actual workload execution.
Although the strategy of caching and over-provisioning resources to enhance responsiveness is intuitive, it raises an important question: To what extent are these performance gains cost-efficient, given the substantial memory overhead incurred?





\textbf{Observation:} \textit{Larger window sizes and keepalive durations increase memory consumption for all evaluated systems by 2.5$\times$ for Kn and 3$\times$ for Kn-Sync, respectively. The utilization target in Kn systems can moderately affect memory consumption.
Across all configurations, allocated memory consistently exceeds the actual demand by instances handling requests, ranging from 2$\times$ to 10$\times$.}



\subsubsection{System Operational Cost}


\begin{figure}
    \centering
    \subfigure[Kn-Sync.]{\includegraphics[width=0.48\linewidth]{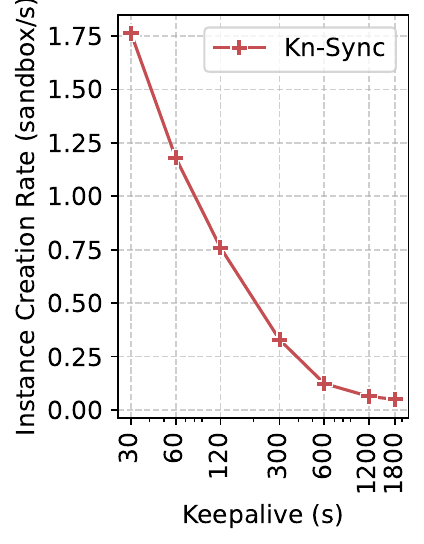}\label{fig:inst_creation_rate:sync}} 
    \subfigure[Kn with different utilization targets.]{\includegraphics[width=0.48\linewidth]{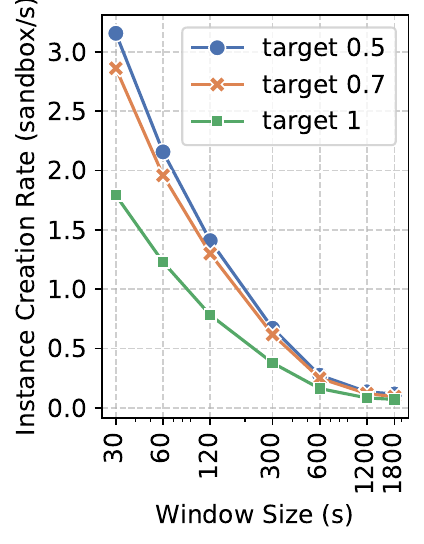}\label{fig:inst_creation_rate:async}} 
    \caption{Comparison of instance creation rates in Kn-Sync and Kn under a realistic workload with different system parameters.
    }
    \label{fig:inst_creation_rate}
\end{figure}






In this section, we evaluate the system operational cost in serverless systems.
We begin by analyzing the instance creation rate under different autoscaling configurations.
Figure~\ref{fig:inst_creation_rate:sync} plots the instance creation rate with different keepalive durations for our synchronous system. We observe that increasing the keepalive duration from 30 seconds to 600 seconds reduces the instance creation rate from 1.8 to 0.12 instances per second. Extending the keepalive to 1800 seconds further reduces the rate to 0.05, representing a 36-fold reduction.
In our evaluated asynchronous system, the instance creation rate is similarly affected by the autoscaling window size, but is also influenced by the utilization target. As is shown in Figure~\ref{fig:inst_creation_rate:async}, with a fixed utilization target of 0.7, increasing the window size from 30 seconds to 1800 seconds reduces the instance creation rate from 2.9 to 0.09 instances per second.
At a fixed window size of 60 seconds, increasing the target value from 0.5 to 1.0 reduces the instance creation rate from 2.2 to 1.2 instances per second, a 45\% decrease.

\begin{figure}
    \centering
    \subfigure[Kn-Sync.]{\includegraphics[width=0.48\linewidth]{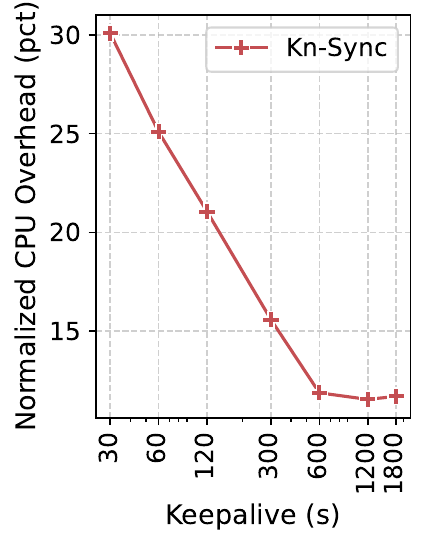}\label{fig:cpu_usage:sync}} 
    \subfigure[Kn with different utilization target.]{\includegraphics[width=0.48\linewidth]{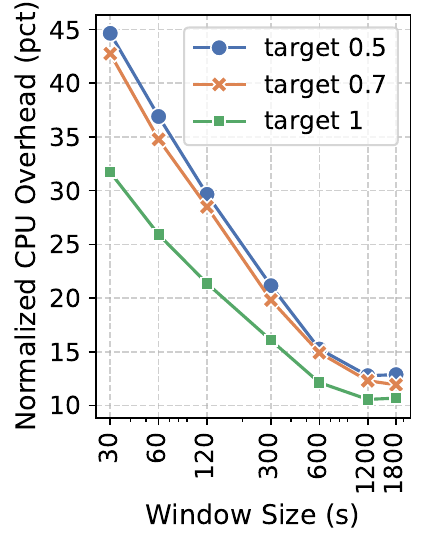}\label{fig:cpu_usage:async}} 
    \caption{Computational overhead in Kn-Sync and Kn under a realistic workload with different system parameters.
    }
    \label{fig:cpu_usage}
\end{figure}



Next, we quantify the amount of computational overhead in the system components. 
Figure~\ref{fig:cpu_usage:sync} presents the measurements of CPU utilization overheads in Knative-Synchronous, normalized to the useful work performed by function instances in handling function invocations. 
We observe that for Knative-Synchronous, the overhead decreases from 30\% to 12\% as the keepalive duration increases from 30 seconds to 600 seconds, and stabilizes beyond 600 seconds.
Figure~\ref{fig:cpu_usage:async} shows the normalized CPU overheads in Knative under varying autoscaling window sizes and utilization targets. For a fixed target of 0.7, the overhead drops from 43\% at a 30-second window to 15\% at 600 seconds, and further extending the window size yields only marginal reductions (12\% at a 1800-second window). At a fixed 30-second window size, decreasing the target from 1.0 to 0.7 results in an 11 percentage point (pp) increase in overhead, while a further decrease from 0.7 to 0.5 leads to only a 2 pp increase.
Comparing Figure~\ref{fig:cpu_usage} with Figure~\ref{fig:inst_creation_rate}, we observe that CPU overhead strongly correlates with the instance creation rate. This suggests that instance lifecycle operations (particularly instance creation) are the most computationally expensive part of cluster management.




\begin{figure}
    \centering
    \includegraphics[width=0.95\linewidth]{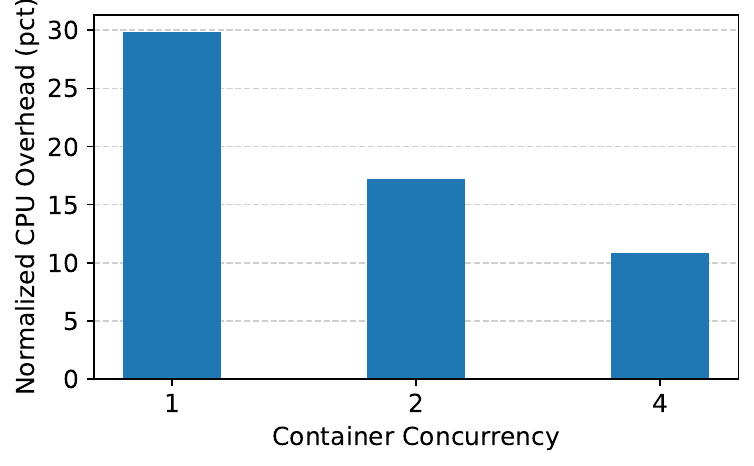}  
    \caption{
    Computational overheads of Kn, utilization target 0.7, window size 60s, under a realistic workload with different container concurrency.
    }
    \label{fig:cpu_usage:cc}
\end{figure}

Lastly, we evaluate the effect of container concurrency (CC) on computational overhead in Kn, with a fixed utilization target of 0.7.
Figure~\ref{fig:cpu_usage:cc} shows that increasing CC from 1 to 4 results in a 3$\times$ reduction in CPU overhead.
This suggests that allowing multiple requests to share a container reduces system operational costs.
\leo{Our evaluation shows that system-level performance metrics do not degrade with higher concurrency capacity (CC) values. However, this result should be considered in light of two methodological limitations. First, increasing CC reduces request-level isolation, potentially introducing interference effects (e.g., resource contention) not captured by our current metrics. Second, the memory footprint is not directly represented. While higher concurrency reduces the number of instances, each instance requires more memory to manage the state of simultaneous requests. Consequently, a lower instance count does not necessarily translate to a net reduction in memory consumption.}


We further characterize the sources of the computational overhead in the cluster. We find that approximately 80\% of the total computational overhead originates from worker nodes, while the remaining 20\% is attributed to the master node. This breakdown suggests that while sandbox management 
(e.g., instance startup, scheduling, and isolation enforcement) 
dominates the computational cost at the worker level, the control plane logic executed on the master node also consumes a nontrivial amount of CPU cycles. \leo{This breakdown doesn't contradict the existing works~\cite{cvetkovic:dirigent} focusing on the control plane as a bottleneck. Computational overheads from worker nodes originate from all nodes in the cluster, whereas control plane overheads are concentrated on a single node, saturating the CPU.}

\textbf{Observation:} \textit{Computational overheads primarily depend on the instance creation rate in the cluster, with the main contributors located on worker nodes (sandbox managers, local agents).}

\subsection{Trade-off between Performance and Cost}
\label{sec:trade-off}



\begin{figure}
    \centering
    \includegraphics[width=0.95\linewidth]{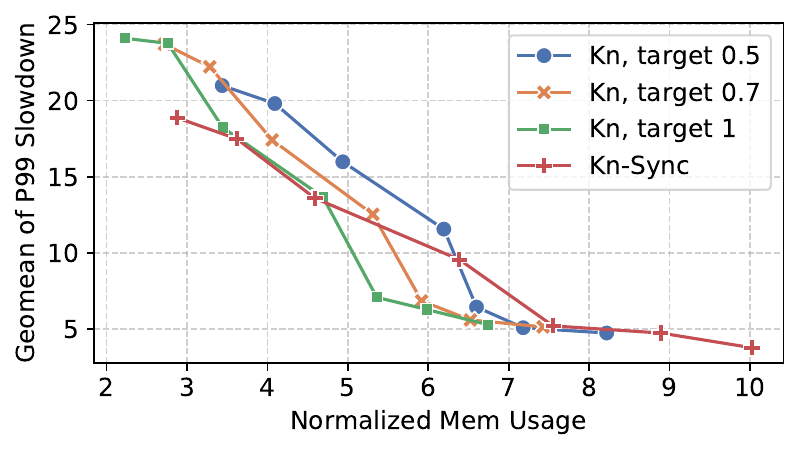}
    \caption{Comparison of Kn-Sync and Kn in cost-performance trade-off space under realistic workload with different system parameters.
    }
    \label{fig:trade-off}
\end{figure}

Now, we combine performance and cost in the same plot  (Figure~\ref{fig:trade-off}) to evaluate how effectively the autoscaling systems utilize memory.
In this plot, the x-axis represents normalized memory usage, while the y-axis captures system performance (slowdown).
Configurations that appear closer to the lower left corner indicate higher efficiency, as they achieve better performance with less memory consumption.

Both evaluated systems
struggle to achieve a slowdown lower than 5, 
regardless of the amount of additional memory. 
This indicates a performance floor, beyond which provisioning more instances yields no tangible benefit.

For Knative, while both target and window size have an impact on the performance, they have different costs for those performance improvements. Out of all targets, the best one is value 1, as it achieves better performance at the same level of memory cost. 

Extra instances created by increasing the window size yield greater performance gains compared to the same number created by lower targets. For example, starting from a configuration with a target value of 1 and a 30-second window, both increasing the window size to 120 seconds and decreasing the target to 0.5 raise the normalized memory usage to approximately 3.4. However, the former results in lower slowdown.


\textbf{Observation:} \textit{The most cost-efficient system for the given performance requirements is Kn, with target values of 1.}




\subsection{Trade-off Evaluation in a Simulated Cluster with a Real-World Control Plane}
\label{sec:big_trade-off}


\begin{figure}
    \centering
    \includegraphics[width=0.95\linewidth]{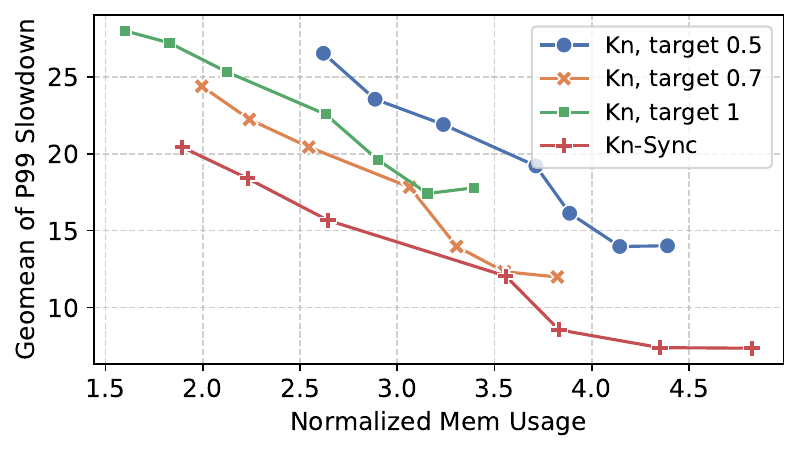}
    \caption{Comparison of Kn-Sync and Kn in cost-performance trade-off space under realistic workload with different system parameters with KWOK.
    }
    \label{fig:trade-off_kwok}
\end{figure}



In this section, we extend our evaluation to a larger scale, using a 2000-function trace with a total of 3.5 million invocations. \leo{This trace requires 10 times more resources to run compared to the 400-function trace used in~\S\ref{sec:trade-off}. We run this trace in a partially simulated environment (see~\S\ref{sec:kwok}) with 50 worker nodes.}
\leo{Memory usage and performance show trends similar to the ones for the smaller trace (\S\ref{sec:trends}) when changing keepalive duration, window size, and utilization target parameters.}

However, we observe different results in the memory-performance trade-off space depicted in Figure~\ref{fig:trade-off_kwok}. The Knative-Synchronous becomes the optimal policy among the evaluated ones, as it can achieve the same or better performance at a lower cost. Another difference is the saturation points for all of the policies. With this workload, policies have different slowdown levels: Knative-Synchronous is saturated at a slowdown of 7, while Knative policies are saturated at slowdowns of 12-18.

These differences in results can be attributed to variations in workload characteristics. Even though the traces are sampled down from full Azure Trace with a methodology that attempts to replicate the original distribution of load between functions, sampled traces won't reproduce it perfectly. This leads to the conclusion that larger, more representative samples are necessary in evaluations. Additionally, a larger trace creates a higher load on the cluster manager, which may impact performance due to the introduction of new bottlenecks.



\textbf{Observation:} \textit{\leo{KWOK-based simulation of worker nodes with real control plane enables large-scale evaluation of serverless systems.} Bigger workloads yield different results in the trade-off space, prompting a more thorough evaluation on a larger scale.}

%% file: sections/5-discussion.tex
\section{Discussion and Future Directions}






In this section, we discuss our key findings and project future research directions towards the serverless systems' performance and cost-efficiency goals.
Our findings reveal a computational overhead that is equivalent to 10-40\% of the CPU cycles consumed by function execution.
This overhead primarily originates from the worker nodes and is attributed to sandbox operations such as instance startup and teardown.
Prior works~\cite{shahrad:serverless, agache:firecracker, li:rund} mainly focus on reducing cold-start latency or sandbox launch time, without thoroughly analyzing the computational costs of sandbox management. Even fewer studies consider the control plane overhead introduced by autoscaling policy and orchestration components~\cite{cvetkovic:dirigent}, which we also find to be non-negligible.
In terms of memory usage, we observe substantial waste due to idle instances, far exceeding the amount required to serve active workloads. 
By contrast, the per-instance memory overhead introduced by modern sandboxing mechanisms such as Firecracker and RunD is minimal~\cite{agache:firecracker, li:rund}. For example, under a configuration with 128MB of memory (the most commonly used memory quota in AWS Lambda~\cite{datadog}), Firecracker incurs a memory overhead of approximately 2\% and RunD less than 15\%. In contrast, idle instances consume 1-9$\times$ more memory than what is used by active instances. This comparison highlights that optimizing scaling policies and idle instance retention strategies offers significantly greater potential for improving memory efficiency than further reducing already minimal sandbox costs.


\begin{table}
\centering
\caption{Impact of Parameter Increase
}
\label{tab:trends}
\begin{tabular}{|l|c|c|c|}
\hline
\textbf{Parameter}              & \textbf{Slowdown} & \textbf{\begin{tabular}[c]{@{}l@{}}Memory\\ usage\end{tabular}} & \textbf{\begin{tabular}[c]{@{}l@{}}Management\\ overhead\end{tabular}} \\ \hline
\hline
\textbf{Keepalive}                & $\Downarrow$        & $\Uparrow$                                                       & $\Downarrow$                                                               \\ \hline
\textbf{Window size}              & $\Downarrow$          & $\Uparrow$                                                        & $\Downarrow$                                                               \\ \hline
\textbf{Utilization target}       & $\Uparrow$          & $\Downarrow$                                                        & $\Downarrow$                                                               \\ \hline
\textbf{Container conc.} & $\approx^{\mathrm{a}}$      & $\Downarrow^{\mathrm{a}}$                                                    & $\Downarrow$                                                               \\ \hline
\multicolumn{4}{l}{$^{\mathrm{a}}$These trends might vary for different applications.}

\end{tabular}
\end{table}

Although the operational costs (both memory and CPU utilization) can be reduced by tuning parameters in the autoscaling policy, we demonstrate that these changes come at the cost of degraded performance or isolation guarantees. We summarize those trends in Table~\ref{tab:trends}. Therefore, it is crucial to design autoscaling strategies that balance efficiency and performance more effectively.




%% file: sections/6-related.tex
\section{Related Work}






\textbf{Realistic serverless traces.} 
Prior work has collected and released several real-world serverless traces, enabling researchers to study workload characteristics and system behavior under production-like conditions.
Shahrad \textit{et al.}~\cite{shahrad:serverless} published the earliest Azure Functions trace, containing millions of invocations with aggregated statistics such as execution durations and memory usage. We use this dataset as the source trace for our workload generation.
Zhang \textit{et al.}~\cite{zhang:faster} provided a finer-grained Azure trace with precise invocation timestamps and durations.
Joosen \textit{et al.}~\cite{joosen:how, joosen:serverless} released traces from Huawei Cloud, capturing cold starts and resource dynamics across multiple data centers. 
Wang \textit{et al.}~\cite{wang:faasnet} published a trace from Alibaba Function Compute, including direct cold start observations and latency measurements.
While these traces are invaluable for characterizing workloads and informing system design, they only offer an external view of system behavior. They do not reveal how internal control planes respond to these workloads. In contrast, our work focuses on replaying real traces against configurable serverless platforms to evaluate autoscaler behavior and platform trade-offs under realistic conditions.



\textbf{Serverless benchmarking frameworks.} 
Prior work includes several benchmarking frameworks and suites designed to analyze serverless platforms.
FaaSDom~\cite{maissen:faasdom}, SeBS~\cite{copik:sebs}, and BeFaaS~\cite{grambow:befaas} provide automated deployment and benchmarking frameworks across multiple runtimes and providers. 
ServerlessBench~\cite{yu:characterizing} and FunctionBench~\cite{kim:functionbench, kim:practical} offer collections of microbenchmarks and real-world workloads to evaluate performance and cost across different cloud platforms.
However, these frameworks primarily emphasize application-level performance or the evaluation of specific test cases, such as function chaining or concurrent cold starts. 
STeLLAR~\cite{ustiugov:stellar} adopts a systems-oriented approach by systematically stressing internal control plane components to analyze their impact on scaling tail latency. 
While scaling latency is also a key concern in our work, STeLLAR focuses on identifying infrastructure-level bottlenecks under synthetic workloads. In contrast, we evaluate how autoscaling strategies and configuration choices impact performance under realistic workloads.
vSwarm-$\mu$~\cite{schall:lukewarm} uses full-system simulation via gem5~\cite{binkert:gem5} to explore hardware-software interactions at the microarchitectural level, whereas our work targets a higher level of abstraction, focusing on the macro-scale behavior and control plane performance. 
In-Vitro~\cite{ustiugov:enabling} proposes a method for processing and sampling data from production traces to construct scalable and configurable workloads. We adopt this approach as the basis for workload generation in our benchmarking framework.
These frameworks provide valuable insights into serverless behavior at different abstraction levels. 
However, prior works have not conducted cross-platform evaluations of autoscaling policies and configuration parameters under consistent and controlled conditions. 
In contrast, we replay real-world workloads to systematically evaluate how different autoscaling strategies and parameter choices affect performance and overhead.


\textbf{Control plane architecture and scheduling policies.}
\leo{Prior work introduced several new control plane architectures and autoscaling approaches.}
IceBreaker~\cite{roy:icebreaker} uses the Fourier transform to predict the invocation concurrency of a function. Based on prediction results, the system decides whether and when to pre-warm the function. 
Atoll~\cite{singhvi:atoll} redesigns the control plane, decoupling the sandbox initialization from the request critical path to eliminate cold start delays and improve tail latency.
Shahrad \textit{et al.}~\cite{shahrad:serverless} analyze a large-scale Azure Functions trace and propose a Hybrid Histogram Policy, which combines long-term historical histograms with short-term recent activity to enable proactive pre-warming. 
Joosen \textit{et al.}~\cite{joosen:how} further confirm that serverless functions exhibit strong periodicity at both individual and aggregate levels, motivating predictive autoscaling based on time-series analysis.
Dirigent~\cite{cvetkovic:dirigent} proposes a novel cluster manager design that significantly improves control plane throughput and latency.
\leo{The works mentioned above lack detailed cost and performance analysis or focus on proxy metrics for performance (e.g., cold start, prediction accuracy).}
Our evaluation methodology supports trace-driven experimentation under realistic load dynamics, enabling direct and fair comparisons of diverse autoscaling and scheduling policies under equivalent conditions.
Another work, PulseNet~\cite{kondrashov:pulsenet}, enhances performance by establishing a dual-track control plane that offers nearly full compatibility with conventional cluster managers and characterizes the cost and performance of the presented system. Our evaluation targets a cross-policy comparison and is more comprehensive, including additional parameters and trend analysis.


%% file: sections/7-conclusion.tex
\section{Conclusion}

In this work, we address key challenges in understanding the performance-cost trade-offs of serverless control plane designs by developing a serverless system that closely approximates the behaviors of commercial providers, such as AWS Lambda and Google Cloud Run, enabling systematic cross-platform evaluation using real production workloads. Our results show that serverless systems incur significant computational overhead (\leo{10-40\% higher CPU utilization}), primarily due to worker node operations, and suffer from substantial memory overprovisioning (2–10$\times$ above active usage), with scaling policy—rather than per-sandbox overheads—emerging as the primary optimization target. We further demonstrate that while tuning autoscaling policies can reduce operational costs, such optimizations often degrade performance, highlighting the need for new cost-efficient autoscaling strategies. By validating our findings at production scale using a hybrid simulation methodology, we bridge the gap between small-scale research and real-world deployments, providing a reproducible framework and actionable insights for future research and practical optimization of serverless platforms at scale.